\documentclass{spie}%
\usepackage{amsfonts}
\usepackage{amsmath}
\usepackage{amssymb}
\usepackage{graphicx}%
\providecommand{\U}[1]{\protect\rule{.1in}{.1in}}

\begin{document}

\title{Photon position eigenvectors lead to complete photon wave mechanics}
\author{Margaret Hawton \skiplinehalf Lakehead Univeristy, Thunder Bay, Canada}
\authorinfo{E-mail: margaret.hawton@lakeheadu.ca, Copyright 2007 Society of Photo-Optical
Engineers and Optical Society of America. This paper was published in SPIE
6664, The Nature of Light: What are Photons?, 2007, pages 666408-1 to 12 and
is made available as an electronic reprint with permission of SPIE and OSA.
One print or electronic copy may be made for personal use only. Systematic or
multiple reproduction, distribution to multiple locations via electric or
other means, duplication of any material in this paper for a fee or for
commercial purposes, or modification of the content of the paper are prohibited.}
\maketitle

\begin{abstract}
We have recently constructed a photon position operator with commuting
components. This was long thought to be impossible, but our position
eigenvectors have a vortex structure like twisted light. Thus they are not
spherically symmetric and the position operator does not transform as a
vector, so that previous non-existence arguments do not apply. We find two
classes of position eigenvectors and obtain photon wave functions by
projection onto the bases of position eigenkets that they define, following
the usual rules of quantum mechanics. The hermitian position operator, r0,
leads to a Landau-Peierls wave function, while field-like eigenvectors of the
nonhermitian position operator and its adjoint lead to a biorthonormal basis.
These two bases are equivalent in the sense that they are related by a
similarity transformation. The eigenvectors of the nonhermitian operators
r$\pm1/2$ lead to a field-potential wave function pair. These field-like
positive frequency wave functions satisfy Maxwell's equations, and thus
justify the supposition that MEs describe single photon wave mechanics. The
expectation value of the number operator is photon density with undetected
photons integrated over, consistent with Feynman's conclusion that the density
of non-interacting particles can be interpreted as probability density.

\end{abstract}

\section{Introduction}

It has been claimed since the early days of quantum mechanics that there is no
position operator for the photon, and thus that a real space wave function
cannot be obtained in the usual way, by projection of the state vector onto a
basis of position eigenvectors. Quite recently we have succeeded in
constructing photon position operators with commuting components whose
transverse eigenvectors form bases of localized states
\cite{HawtonPO,HawtonBaylisPO,HawtonBaylisAM}. These Hermitian and
pseudo-Hermitian bases can be used to derive photon wave functions in the
usual way, leading to a real space description of the photon that is
compatible with all of the usual rules of quantum mechanics
\cite{HawtonWaveMechanics07}. It is this new photon wave mechanics that will
be discussed here.

\section{m%
$>$%
0 wave mechanics}

In the standard formulation of quantum mechanics an observable is described by
an operator, $\widehat{o},$ whose eigenvalues, $o_{n},$ are the possible
results of a measurement of that observable. If the state vector is expanded
in its complete set of eigenvectors, $\left\{  \left\vert \phi_{n}%
\right\rangle \right\}  ,$ the squares of the coefficients in this expansion
are the probabilities for the corresponding eigenvalue to be measured, that is%
\begin{align*}
\widehat{o}\left\vert \psi_{n}\right\rangle  &  =o_{n}\left\vert \psi
_{n}\right\rangle ,\\
\left\vert \Psi\right\rangle  &  =\sum_{n}c_{n}\left\vert \psi_{n}%
\right\rangle ,\\
\left\vert c_{n}\right\vert ^{2}  &  =\text{probability to get }o_{n}.
\end{align*}
If the observable is position, the spectrum of eigenvalues becomes continuous
so that%
\begin{align}
\widehat{\mathbf{r}}\left\vert \phi_{\mathbf{r}^{\prime}}\right\rangle  &
=\mathbf{r}^{\prime}\left\vert \phi_{\mathbf{r}^{\prime}}\right\rangle
,\label{rOpEq}\\
\left\vert \Psi\right\rangle  &  =\int d^{3}r^{\prime}c\left(  \mathbf{r}%
^{\prime}\right)  \left\vert \phi_{\mathbf{r}^{\prime}}\right\rangle
,\nonumber\\
\left\vert c_{\mathbf{r}^{\prime}}\right\vert ^{2}  &  =\text{probability
density to find the particle at }\mathbf{r}^{\prime}.\nonumber
\end{align}
In real space the position operator of a massive particle such as an electron
is $\widehat{\mathbf{r}}=\mathbf{r}$ and the position eigenvectors are the
$3$-dimensional Dirac $\delta$-functions,
\[
\psi_{\mathbf{r}^{\prime}}\left(  \mathbf{r}\right)  =\delta^{3}\left(
\mathbf{r}-\mathbf{r}^{\prime}\right)  ,
\]
localized at $\mathbf{r}^{\prime}$. The $\mathbf{k}$-space description of
these localized states is their inverse Fourier transform,
\begin{align*}
\psi_{\mathbf{r}^{\prime}}\left(  \mathbf{k}\right)   &  =\int d^{3}%
r\delta^{3}\left(  \mathbf{r}-\mathbf{r}^{\prime}\right)  \frac{\exp\left(
-i\mathbf{k\cdot r}\right)  }{\sqrt{V}}\\
&  =\frac{\exp\left(  -i\mathbf{k\cdot r}^{\prime}\right)  }{\sqrt{V}},
\end{align*}
while the position operator becomes%
\begin{equation}
\widehat{\mathbf{r}}=i\nabla. \label{PositionOp}%
\end{equation}
Here $\nabla$ is the $\mathbf{k}$-space gradient operator whose $j^{th}$
Cartesian component is $\nabla_{j}=\partial/\partial k_{j}$.

\section{Glauber wave function}

In spite of the widely held belief that there is no photon position operator
and hence no first quantized description of the photon, many quantum optics
researchers consider a photon wave function to be highly desirable. This was
the subject of a number of papers at the 2005 SPIE conference on The Photon
and in a special issue of Optics and Photonics News \cite{ThePhoton}. A wave
function can be introduced through Glauber photodetection theory
\cite{Glauber}. The Glauber probability per unit time that a photon will be
absorbed by a negligibly small detector at point $\mathbf{r}$ and time $t$ for
an electromagnetic field in state $\left\vert \Psi\right\rangle $\ is%
\begin{equation}
\frac{dn_{G}}{dt}\propto\left\langle \Psi\left\vert \widehat{\mathbf{E}}%
^{(-)}\left(  \mathbf{r},t\right)  \cdot\widehat{\mathbf{E}}^{(+)}\left(
\mathbf{r},t\right)  \right\vert \Psi\right\rangle . \label{Glauber}%
\end{equation}
The $1$-photon wave function is then defined as the probability amplitude
\cite{ScullyBook,Sipe}%
\begin{equation}
\mathbf{\Psi}^{(1/2)}\left(  \mathbf{r},t\right)  =\left\langle 0\left\vert
\widehat{\mathbf{E}}^{(+)}\left(  \mathbf{r},t\right)  \right\vert
\Psi\right\rangle . \label{GlauberPsi}%
\end{equation}
The superscript $1/2$ used here denotes the quantum electrodynamics (QED)
$\left(  \omega_{k}\right)  ^{1/2}$ dependence of the field operator%
\begin{equation}
\widehat{\mathbf{E}}^{(+)}\left(  \mathbf{r},t\right)  \equiv\sum_{\mathbf{k}%
}\mathcal{C}\left(  \omega_{k}\right)  ^{1/2}\mathbf{e}_{\mathbf{k},\sigma
}^{(\chi)}a_{\mathbf{k},\sigma}\frac{e^{i\mathbf{k\cdot r}-i\omega_{k}t}%
}{\sqrt{V}} \label{Eop}%
\end{equation}
where $\omega_{k}=kc$ in free space. This is consistent with the QED based
interpretation that a mode with frequency $\omega_{k}$ has energy $\hbar
\omega_{k}$. Since the square of the field must give energy density, terms
must be weighted by the factor $\sqrt{\omega_{k}}$. The operator
$a_{\mathbf{k},\sigma}$ annihilates a photon with wave vector $\mathbf{k}$ and
polarization $\sigma$ and $\mathcal{C}\equiv\sqrt{\hbar/2\epsilon_{0}V}$. This
wave function is useful, but its square is not a probability density, and it
cannot be incorporated into the usual quantum mechanical framework.

There are a number of other candidates for photon wave function in the
literature \cite{BB}. In particular, the Landau-Peierls (LP) wave function
\cite{LandauPeierls}, $\mathbf{\Psi}^{(0)}\left(  \mathbf{r},t\right)  ,$
omits the $\sqrt{\omega_{k}}$ dependence. Its absolute value squared,
$n^{(0)}\left(  \mathbf{r},t\right)  =\left\vert \mathbf{\Psi}^{(0)}\left(
\mathbf{r},t\right)  \right\vert ^{2},$ is then a local and positive definite
photon number density. This wave function has the disadvantage that its
relationship to current is nonlocal. Equations of motion and probability
density based on $\widehat{\mathbf{\Psi}}^{(0)}$ was investigated by Mandel
and Cook \cite{Cook}. To allow for both LP and field-like wave function we
will consider a frequency dependence of the form $\left(  \omega_{k}\right)
^{\alpha}$ here.

\section{Spin and helicity}

The photon helicity is the component of spin in the direction of propagation,
that is in the $\mathbf{k}$-direction. We will work in $\mathbf{k}$-space,
which is equivalent to momentum space. The localized basis states that we will
use here have definite helicity, but they do not have definite spin or orbital
angular momentum (AM) along any fixed direction in space. Since it is
important to understand these distinctions, we will first consider the
$z$-components of orbital AM, spin, and helicity in units of $\hbar$,
\begin{align}
\widehat{L}_{z}  &  =i\frac{\partial}{\partial\phi},\label{OAM}\\
S_{z}  &  =\left(
\begin{array}
[c]{ccc}%
0 & -i & 0\\
i & 0 & 0\\
0 & 0 & 0
\end{array}
\right)  ,\label{Spin}\\
S_{\mathbf{k}}  &  =\widehat{\mathbf{\mathbf{k}}}\mathbf{\cdot S.}
\label{Helicity}%
\end{align}
The Cartesian components of the spin $1$ operator are the $3\times3$ matrices
$\left(  S_{i}\right)  _{jk}=-i\epsilon_{ijk}$, with its $z$-component being
given explicitly in (\ref{Spin}). The eigenvectors of $\widehat{L}_{z}$ are
$\exp\left(  il_{z}\phi\right)  /\sqrt{2\pi}$ with eigenvalue $\hbar l_{z},$
while those of \ $\widehat{S}_{z}$ are $\mathbf{e}_{s_{z}}=\left(
\widehat{\mathbf{x}}+is_{z}\widehat{\mathbf{y}}\right)  /\sqrt{2}$ with
eigenvalue $s_{z}=\pm1$ and $\mathbf{e}_{0}=\widehat{\mathbf{z}}$ with
eigenvalue $s_{z}=0.$ The unitary transformation%
\begin{equation}
D=\exp\left(  -iS_{z}\phi\right)  \exp\left(  -iS_{y}\theta\right)  \label{D}%
\end{equation}
rotates the Cartesian unit vectors, $\left\{  \widehat{\mathbf{x}}%
,\widehat{\mathbf{y}},\widehat{\mathbf{z}}\right\}  ,$ to the spherical polar
unit vectors $\left\{  \widehat{\mathbf{\theta}},\widehat{\mathbf{\phi}%
},\widehat{\mathbf{k}}\right\}  .$ It transforms $S_{z}$ to the helicity
operator $S_{\mathbf{k}}=DS_{z}D^{-1}$ and the $S_{z}$ eigenvectors, $\left(
\widehat{\mathbf{x}}+is_{z}\widehat{\mathbf{y}}\right)  /\sqrt{2},$ to
$\mathbf{e}_{\mathbf{k},\sigma}^{(0)}=D\mathbf{e}_{s_{z}}$ where
\begin{equation}
\mathbf{e}_{\mathbf{k},\sigma}^{(0)}=\left(  \widehat{\mathbf{\theta}}%
+i\sigma\widehat{\mathbf{\phi}}\right)  /\sqrt{2}. \label{e0}%
\end{equation}
This preserves eigenvalues so that unit vectors $\mathbf{e}_{\mathbf{k}%
,\sigma}^{(0)}$ have definite helicity $\sigma=\pm1,$ while $\mathbf{e}%
_{\mathbf{k},0}^{(0)}=\widehat{\mathbf{k}}$ has eigenvalue $0$. The unit
vectors are no longer fixed, since $\widehat{\mathbf{\theta}}$ and
$\widehat{\mathbf{\phi}}$ depend on $\widehat{\mathbf{k}}.$

\section{Position operator}

In the absence of free charge the displacement and magnetic induction vectors
satisfy $\nabla\cdot\mathbf{D}=0$ and $\nabla\cdot\mathbf{B}=0,$ that is these
fields are transverse. In the Coulomb gauge the vector potential is also
transverse.\ In $\mathbf{k}$-space space this condition reduces to
$\mathbf{k}\cdot\mathbf{D}=0$ and $\mathbf{k}\cdot\mathbf{B}=0,$ and the
spherical polar unit vectors\ $\widehat{\mathbf{\theta}}$ and $\widehat
{\mathbf{\phi}}$ provide a basis for the description of these transverse
vectors. We will use the equivalent $\mathbf{e}_{\mathbf{k},\sigma}^{(0)}$
basis here, since eigenvectors of the helicity operator provide us with a
complete set of commuting observables (CSCO), namely helicity and the
components of the position \ operator. Transverse wave functions of the form
\cite{HawtonPO}
\begin{equation}
\psi_{\mathbf{r},\sigma,j}^{(\alpha)}\left(  \mathbf{k}\right)  =\left(
\omega_{k}\right)  ^{\alpha}e_{\mathbf{k},\sigma,j}^{(\chi)}\exp\left(
-i\mathbf{k\cdot r}\right)  /\sqrt{V} \label{PositionEigenvectors}%
\end{equation}
describe a photon with helicity $\sigma$ located at position $\mathbf{r}.$
Here subscripts denote eigenvalues and Cartesian components of the vectors
$\mathbf{\psi}$ and $\mathbf{e,}$ and Cartesian components are used where it
is necessary to avoid confusing vector notation. The parameter $\alpha$
describes electric and/or magnetic fields if $\alpha=1/2,$ the vector
potential if $\alpha=-1/2,$ or LP wave functions if $\alpha=0.$ Periodic
boundary conditions in a finite volume are used here to simplify the notation,
and the limit as $V\rightarrow\infty$ can be taken to calculate derivatives
and perform sums.

If the wave function (\ref{PositionEigenvectors}) is a position eigenvector it
should satisfy the eigenvector equation (\ref{rOpEq}), that is%
\begin{equation}
\widehat{\mathbf{r}}^{(\alpha)}\psi_{\mathbf{r},\sigma,j}^{(\alpha)}\left(
\mathbf{k}\right)  =\mathbf{r}\psi_{\mathbf{r},\sigma,j}^{(\alpha)}\left(
\mathbf{k}\right)  \label{EvectEq}%
\end{equation}
where $\widehat{\mathbf{r}}^{(\alpha)}$ is the $\mathbf{k}$-space
representation of the position operator and its eigenvalues, $\mathbf{r,}$ can
be interpreted as photon position. The operator that satisfies (\ref{EvectEq})
is \cite{HawtonPO}
\begin{equation}
\widehat{\mathbf{r}}^{(\alpha)}=iI\nabla-iI\alpha\widehat{\mathbf{k}%
}\mathbf{/}k+\widehat{\mathbf{k}}\mathbf{\times S}/k+S_{\mathbf{k}}%
\widehat{\mathbf{\phi}}\cot\theta/k \label{rop}%
\end{equation}
where $I$ is a $3\times3$ unit matrix. Eq. (\ref{rop}) looks cumbersome, but
it has a simple interpretation. The position information is contained in the
phase factor, $\exp\left(  -i\mathbf{k\cdot r}\right)  $. The position
operator (\ref{rop}) can be rewritten as \cite{HawtonBaylisPO}%
\[
\widehat{\mathbf{r}}^{(\alpha)}=D\left[  \left(  \omega_{k}\right)  ^{\alpha
}i\nabla\left(  \omega_{k}\right)  ^{-\alpha}\right]  D^{-1}.
\]
The interpretation of this is that, starting from a wave vector parallel to
$\widehat{\mathbf{z}}$ and transverse unit vectors $\widehat{\mathbf{x}}$ and
$\widehat{\mathbf{y}},$ $D$ rotates $\mathbf{k}$ from the $z$-axis to an
orientation described by the angles $\theta$ and $\phi,$ while at the same
time rotating the transverse vectors to $\widehat{\mathbf{\theta}}$ and
$\widehat{\mathbf{\phi}}$ to give (\ref{e0}). The factor $\left(  \omega
_{k}\right)  ^{-\alpha}$ removes the $\left(  \omega_{k}\right)  ^{\alpha}%
\ $dependence in (\ref{PositionEigenvectors}). For example, when
$\widehat{\mathbf{r}}^{(1/2)}$ acts on a transverse field parallel to
$\widehat{\mathbf{\phi}},$ $D^{-1}$ rotates it to $\widehat{\mathbf{y}}$ and
divides it by $\sqrt{\omega_{k}},$ leaving only the phase factor which it
operates on it with the usual $\mathbf{k}$-space position operator, $i\nabla$
(\ref{PositionEigenvectors}). The process is then reversed by multiplying by
$\sqrt{\omega_{k}}$ and using $D$ to rotate the eigenvector back to its
original transverse orientation. This allows $\widehat{\mathbf{r}}^{(1/2)}$ to
extract the position of the photon from $\exp\left(  -i\mathbf{k\cdot
r}\right)  $ while ignoring other factors.

The position operator can be generalized to allow for rotation about
$\mathbf{k}$ through the Euler angle $\chi\left(  \theta,\phi\right)  \ $to
give the most general transverse basis \cite{HawtonBaylisPO},
\begin{equation}
\mathbf{e}_{\mathbf{k},\sigma}^{\left(  \chi\right)  }=e^{-i\sigma\chi
}\mathbf{e}_{\mathbf{k},\sigma}^{(0)}. \label{e_chi}%
\end{equation}
The quantum numbers $\left\{  \mathbf{r},\sigma\right\}  $ index the basis
states of the CSCO for a given $\chi\left(  \theta,\phi\right)  $. The
$z$-axis can be selected for convenience and the choice $\chi=-m\phi$ then
gives \cite{HawtonBaylisAM}%
\begin{align}
\mathbf{e}_{\mathbf{k},\sigma}^{\left(  -m\phi\right)  }  &  =\frac
{\widehat{\mathbf{x}}-i\widehat{\mathbf{y}}}{2\sqrt{2}}\left(  \cos
\theta-\sigma\right)  e^{i\left(  m\sigma+1\right)  \phi}-\frac{\widehat
{\mathbf{z}}}{\sqrt{2}}\sin\theta e^{im\sigma\phi}\nonumber\\
&  +\frac{\widehat{\mathbf{x}}+i\widehat{\mathbf{y}}}{2\sqrt{2}}\left(
\cos\theta+\sigma\right)  e^{i\left(  m\sigma-1\right)  \phi}. \label{em}%
\end{align}
For example, $\chi=-\phi$ ($m=1$) rotates $\widehat{\mathbf{\theta}}$ and
$\widehat{\mathbf{\phi}}$ back to the $x$ and $y$ axes to give unit vectors
that approach $\left(  \widehat{\mathbf{x}}+i\sigma\widehat{\mathbf{y}%
}\right)  /\sqrt{2}$ in the $\theta\rightarrow0$ paraxial limit that describes
the optical beams commonly available in the laboratory.

\section{Inner-product and similarity transformation}

The $\alpha=0$ position operator is Hermitian but, for $\alpha=\pm1/2,$
$\widehat{\mathbf{r}}$ can be Hermitian or pseudo-Hermitian depending on the
choice of inner-product, as will be argued next. The LP inner-product is%
\begin{equation}
\left\langle \mathbf{\Psi}^{(0)}|\widetilde{\mathbf{\Psi}}^{(0)}\right\rangle
=\sum_{\mathbf{k},j}\Psi{}_{j}^{(0)\ast}\left(  \mathbf{k}\right)
\widetilde{\Psi_{j}}^{(0)}\left(  \mathbf{k}\right)  , \label{ScalarProduct0}%
\end{equation}
and it can be verified using integration by parts that $\left\langle
\mathbf{\Psi}^{(0)}|\widehat{\mathbf{r}}^{(0)}\widetilde{\mathbf{\Psi}}%
^{(0)}\right\rangle =\left\langle \widehat{\mathbf{r}}^{(0)}\mathbf{\Psi
}^{(0)}|\widetilde{\mathbf{\Psi}}^{(0)}\right\rangle ,$ which implies that
$\widehat{\mathbf{r}}^{(0)}$ is Hermitian. Alternatively, the usual rules for
calculating adjoints give%
\begin{align*}
\widehat{\mathbf{r}}^{(0)\dagger}  &  =\left(  iI\nabla+\widehat{\mathbf{k}%
}\mathbf{\times S}/k+S_{\mathbf{k}}\widehat{\mathbf{\phi}}\cot\theta/k\right)
^{\dagger}\\
&  =iI\nabla+\widehat{\mathbf{k}}\mathbf{\times S}/k+S_{\mathbf{k}}%
\widehat{\mathbf{\phi}}\cot\theta/k=\widehat{\mathbf{r}}^{(0)}%
\end{align*}
which confirms that $\widehat{\mathbf{r}}^{(0)}$\ is Hermitian. However,
application of the same rules to the $\alpha=1/2$ case gives%
\begin{align*}
\widehat{\mathbf{r}}^{(1/2)\dagger}  &  =\left(  iI\nabla-\frac{1}{2}%
iI\alpha\widehat{\mathbf{k}}\mathbf{/}k+\widehat{\mathbf{k}}\mathbf{\times
S}/k+S_{\mathbf{k}}\widehat{\mathbf{\phi}}\cot\theta/k\right)  ^{\dagger}\\
&  =iI\nabla+\frac{1}{2}iI\alpha\widehat{\mathbf{k}}\mathbf{/}k+\widehat
{\mathbf{k}}\mathbf{\times S}/k+S_{\mathbf{k}}\widehat{\mathbf{\phi}}%
\cot\theta/k\\
&  =\widehat{\mathbf{r}}^{(-1/2)}.
\end{align*}
While a field theory inner-product $\left\langle \mathbf{\Psi}^{(1/2)}%
|\widetilde{\mathbf{\Psi}}^{(1/2)}\right\rangle =\sum_{\mathbf{k},j}k^{-1}%
\Psi^{(1/2)\ast}\left(  \mathbf{k}\right)  \widetilde{\Psi}^{(1/2)}\left(
\mathbf{k}\right)  $ does make $\widehat{\mathbf{r}}^{(1/2)}$ Hermitian, this
results in a nonlocal photon number density. An alternative approach is to
define the inner-product%
\begin{equation}
\left\langle \mathbf{\Psi}^{(1/2)}|\widetilde{\mathbf{\Psi}}^{(-1/2)}%
\right\rangle =\sum_{\mathbf{k},j}\Psi_{j}^{(1/2)\ast}\left(  \mathbf{k}%
\right)  \widetilde{\Psi_{j}}^{(-1/2)}\left(  \mathbf{k}\right)
\label{ScalarProduct1/2}%
\end{equation}
involving biorthonormal field-potential pairs. The operator pair
$\widehat{\mathbf{r}}^{(1/2)}$ and $\widehat{\mathbf{r}}^{(-1/2)}$ can then be
called pseudo-Hermitian \cite{Mostaf}.

The LP and field-potential bases are related by a similarity transformation.
In linear algebra, a similarity transformation, $\rho,$ of a matrix $O$ with
eigenvectors, $\ f_{n},$ is of the form $O\rightarrow\rho O\rho^{-1}$ and
$f_{n}\rightarrow\rho f_{n}.$ Since $Of_{n}=\lambda f_{n}$ implies $\left(
\rho O\rho^{-1}\right)  \left(  \rho f_{n}\right)  =\lambda\left(  \rho
f_{n}\right)  ,$ this transformation preserves the eigenvalues, $\lambda.$ If
$\rho^{-1}=\rho^{\dagger}$ this similarity transformation is unitary, as is
usual in quantum mechanics. The transformation relating the LP and field-like
bases,
\begin{align}
\widehat{\mathbf{r}}^{(1/2)}  &  =\omega_{k}^{1/2}\widehat{\mathbf{r}}%
^{(0)}\omega_{k}^{-1/2},\label{Similarity}\\
\mathbf{\psi}_{\mathbf{r},\sigma}^{(1/2)}\left(  \mathbf{k}\right)   &
=\omega_{k}^{1/2}\mathbf{\psi}_{\mathbf{r},\sigma}^{(0)}\left(  \mathbf{k}%
\right)  ,\nonumber\\
\mathbf{\psi}_{\mathbf{r},\sigma}^{(-1/2)}\left(  \mathbf{k}\right)   &
=\omega_{k}^{-1/2}\mathbf{\psi}_{\mathbf{r},\sigma}^{(0)}\left(
\mathbf{k}\right)  ,\nonumber
\end{align}
is also a similarity transformation that preserves eigenvalues. Thus the fact
that the eigenvalues of the Hermitian operator $\widehat{\mathbf{r}}^{(0)}$
are real thus implies that the eigenvalues of $\widehat{\mathbf{r}}^{(\pm
1/2)}$ are also real. The inner-product (\ref{ScalarProduct1/2}) is equal to
(\ref{ScalarProduct0}), so the inner-product is also preserved by the
similarity transformation. The transformation (\ref{Similarity}) is not
unitary and the operator $\widehat{\mathbf{r}}^{(1/2)}$ is not Hermitian. This
is outsider the scope of the usual quantum mechanical formalism, but there has
been consider work done on pseudo-Hermitian Hamiltonians in recent years
\cite{Mostaf}.

\section{Position eigenvectors}

We can define operators that annihilate or create a photon at position
$\mathbf{r}$ and time $t$ as
\begin{align}
\widehat{\psi}_{\mathbf{r},\sigma,j}^{(\alpha)}\left(  t\right)   &
\equiv\sum_{\mathbf{k}}\left(  \omega_{k}\right)  ^{\alpha}e_{\mathbf{k}%
,\sigma,j}^{(\chi)}a_{\mathbf{k},\sigma}\frac{e^{i\mathbf{k\cdot r}%
-i\omega_{k}t}}{\sqrt{V}},\label{Annihilation}\\
\widehat{\psi}_{\mathbf{r},\sigma,j}^{(\alpha)\dagger}\left(  t\right)   &
\equiv\sum_{\mathbf{k}}\left(  \omega_{k}\right)  ^{\alpha}e_{\mathbf{k}%
,\sigma,j}^{(\chi)\ast}a_{\mathbf{k},\sigma}^{\dagger}\frac
{e^{-i\mathbf{k\cdot r}+i\omega_{k}t}}{\sqrt{V}}. \label{Creation}%
\end{align}
These are proportional to field operators if $\alpha=1/2$ and to the vector
potential operator if $\alpha=-1/2.$ They satisfy the equal time commutation
relations%
\begin{equation}
\sum_{j}\left[  \widehat{\psi}_{\mathbf{r},\sigma,j}^{(-\alpha)}\left(
t\right)  ,\widehat{\psi}_{\mathbf{r}^{\prime},\sigma^{\prime},j}%
^{(\alpha)\dagger}\left(  t\right)  \right]  =\delta_{\sigma,\sigma^{\prime}%
}\delta^{3}\left(  \mathbf{r}-\mathbf{r}^{\prime}\right)  ,
\label{Commutation}%
\end{equation}
analogous to the displacement-potential commutation relation in QED. In terms
of these operators the $1$-photon localized basis states are%
\begin{equation}
\left\vert \mathbf{\psi}_{\mathbf{r},\sigma}^{(\alpha)}\left(  t\right)
\right\rangle =\widehat{\mathbf{\psi}}_{\mathbf{r},\sigma}^{(\alpha)\dagger
}\left(  t\right)  \left\vert 0\right\rangle . \label{1PhotonEvect}%
\end{equation}
If these position eigenvectors are projected onto the orthonormal
momentum-helicity basis states $\left\vert \mathbf{k},\sigma\right\rangle $,
their $\mathbf{k}$-space representation is found to be%
\begin{align}
\psi_{\mathbf{r},\sigma,j}^{(\alpha)}\left(  \mathbf{k},t\right)   &
=\left\langle \mathbf{k},\sigma\left\vert \widehat{\mathbf{\psi}}%
_{\mathbf{r},\sigma}^{(\alpha)\dagger}\left(  t\right)  \right\vert
0\right\rangle =\left\langle \mathbf{k},\sigma\right\vert \left(
\sum_{\mathbf{k}^{\prime}}\left(  \omega_{k^{\prime}}\right)  ^{\alpha
}e_{\mathbf{k}^{\prime},\sigma,j}^{(\chi)\ast}\frac{e^{-i\mathbf{k}^{\prime
}\mathbf{\cdot r}+i\omega_{k}t}}{\sqrt{V}}\left\vert \mathbf{k}^{\prime
},\sigma\right\rangle \right) \nonumber\\
&  =\left(  \omega_{k}\right)  ^{\alpha}e_{\mathbf{k},\sigma,j}^{(\chi)}%
\frac{e^{i\mathbf{k\cdot r}-i\omega_{k}t}}{\sqrt{V}}. \label{rBasis}%
\end{align}
This is the Heisenberg picture (HP) equivalent of Eq.
(\ref{PositionEigenvectors}).

\section{Wave function}

A coordinate space wave function is the projection of the state vector onto a
basis of position eigenvectors. A QED state vector expanded in the
number-momentum-helicity basis is
\begin{equation}
\left\vert \Psi\right\rangle =c_{0}\left\vert 0\right\rangle +\sum
_{\mathbf{k},\sigma}c_{\mathbf{k},\sigma}a_{\mathbf{k},\sigma}^{\dagger
}\left\vert 0\right\rangle +\frac{1}{2!}\sum_{\mathbf{k},\sigma;\mathbf{k}%
^{\prime},\sigma^{\prime}}\sqrt{\mathcal{N}_{\mathbf{k},\sigma;\mathbf{k}%
^{\prime},\sigma^{\prime}}}c_{\mathbf{k},\sigma;\mathbf{k}^{\prime}%
,\sigma^{\prime}}a_{\mathbf{k},\sigma}^{\dagger}a_{\mathbf{k}^{\prime}%
,\sigma^{\prime}}^{\dagger}\left\vert 0\right\rangle +.. \label{StateVector}%
\end{equation}
where $c_{0}=\left\langle 0|\Psi\right\rangle ,$ $c_{\mathbf{k},\sigma}%
\equiv\left\langle 0\left\vert a_{\mathbf{k},\sigma}\right\vert \Psi
\right\rangle ,$ $c_{\mathbf{k},\sigma;\mathbf{k}^{\prime},\sigma^{\prime}%
}\equiv c_{\mathbf{k}^{\prime},\sigma^{\prime};\mathbf{k},\sigma}=\left\langle
0\left\vert a_{\mathbf{k},\sigma}a_{\mathbf{k}^{\prime},\sigma^{\prime}%
}\right\vert \Psi\right\rangle ,$ and $\mathcal{N}_{\mathbf{k},\sigma
;\mathbf{k}^{\prime},\sigma^{\prime}}=1+\delta_{\mathbf{k},\mathbf{k}^{\prime
}}\delta_{\sigma,\sigma^{\prime}}$. Division by $2!$ corrects for identical
states obtained when the $\left\{  \mathbf{k},\sigma\right\}  $ subscripts are
permuted while $\sqrt{\mathcal{N}}/2$ normalizes doubly occupied states.

The $1$-photon wave function is found by projecting (\ref{StateVector}) onto
(\ref{1PhotonEvect}) to give
\begin{align}
\mathbf{\Psi}_{\sigma}^{(\alpha)}\left(  \mathbf{r},t\right)   &
=\left\langle \mathbf{\psi}_{\mathbf{r},\sigma}^{(\alpha)}|\mathbf{\Psi
}\right\rangle \nonumber\\
&  =\sum_{\mathbf{k}}c_{\mathbf{k},\sigma}\mathbf{e}_{\mathbf{k,}\sigma
}^{(\chi)}\left(  \omega_{k}\right)  ^{\alpha}\frac{e^{i\mathbf{k\cdot
r}-i\omega_{k}t}}{\sqrt{V}}, \label{1PhotonPsi}%
\end{align}
where we are interested in the cases $\alpha=0$ (LP) and $\alpha=\pm1/2$
(fields and vector potential). It can be verified that these wave functions
satisfy%
\begin{equation}
i\frac{\partial\mathbf{\Psi}_{\sigma}^{(-1/2)}\left(  \mathbf{r},t\right)
}{\partial t}=\mathbf{\Psi}_{\sigma}^{(1/2)}\left(  \mathbf{r,}t\right)  ,
\label{FieldPotentialWaveEq}%
\end{equation}
which is analogous to $\mathbf{E=-\partial A/\partial}t$ in the Coulomb gauge
in free space. Using $i\widehat{\mathbf{k}}\times\mathbf{e}_{\mathbf{k,}%
\sigma}^{(0)}=\sigma\mathbf{e}_{\mathbf{k,}\sigma}^{(0)}$ with $\mathbf{e}%
_{\mathbf{k,}\sigma}^{(0)}$ given by (\ref{e0}), it follows that%
\begin{equation}
i\frac{\partial\mathbf{\Psi}_{\sigma}^{(\alpha)}\left(  \mathbf{r},t\right)
}{\partial t}=\sigma c\nabla\times\mathbf{\Psi}_{\sigma}^{(\alpha)}\left(
\mathbf{r},t\right)  \label{WaveEquation}%
\end{equation}
for any $\alpha.$

To obtain the $2$-photon wave function we can project $\left\vert
\Psi\right\rangle $ onto the $2$-photon real space basis
\begin{equation}
\left\vert \psi_{\mathbf{r},\sigma,j}\left(  t\right)  ,\psi_{\mathbf{r}%
^{\prime},\sigma^{\prime},j^{\prime}}\left(  t^{\prime}\right)  \right\rangle
=\widehat{\psi}_{\mathbf{r},\sigma,j}^{(\alpha)\dagger}\left(  t\right)
\widehat{\psi}_{\mathbf{r}^{\prime},\sigma^{\prime},j^{\prime}}^{(\alpha
)\dagger}\left(  t^{\prime}\right)  \left\vert 0\right\rangle \nonumber
\end{equation}
as%
\begin{equation}
\Psi_{\sigma,\sigma^{\prime};j,j^{\prime}}^{(\alpha)}\left(  \mathbf{r}%
,\mathbf{r}^{\prime};t,t^{\prime}\right)  =\left\langle 0\left\vert
\widehat{\psi}_{\mathbf{r},\sigma,j}^{(\alpha)}\left(  t\right)  \widehat
{\psi}_{\mathbf{r}^{\prime},\sigma^{\prime},j^{\prime}}^{(\alpha)}\left(
t^{\prime}\right)  \right\vert \Psi\right\rangle . \label{2Photon}%
\end{equation}
Use of Eq. (\ref{Creation}) and $\left[  a_{\mathbf{k},\sigma},a_{\mathbf{k}%
^{\prime},\sigma^{\prime}}^{\dagger}\right]  =\delta_{\mathbf{k}%
,\mathbf{k}^{\prime}}\delta_{\sigma,\sigma^{\prime}}$ to evaluate
$\left\langle 0\right\vert \widehat{\psi}_{\mathbf{r},\sigma,j}^{(\alpha
)}\left(  t\right)  \widehat{\psi}_{\mathbf{r}^{\prime},\sigma^{\prime
},j^{\prime}}^{(\alpha)}\left(  t^{\prime}\right)  a_{\mathbf{k}%
,\sigma^{\prime\prime}}^{\dagger}a_{\mathbf{k}^{\prime},\sigma^{\prime
\prime\prime}}^{\dagger}\left\vert 0\right\rangle $ then gives
\begin{align}
\Psi_{\sigma,\sigma^{\prime};j,j^{\prime}}^{(\alpha)}\left(  \mathbf{r}%
,\mathbf{r}^{\prime};t,t^{\prime}\right)   &  =\frac{1}{2!V}\sum
_{\mathbf{k},\sigma;\mathbf{k}^{\prime},\sigma^{\prime}}\sqrt{\mathcal{N}%
_{\mathbf{k},\sigma;\mathbf{k}^{\prime},\sigma^{\prime}}}c_{\mathbf{k}%
,\sigma;\mathbf{k}^{\prime},\sigma^{\prime}}\left(  \omega_{k}\omega
_{k^{\prime}}\right)  ^{\alpha}\label{2PhotonWaveFtn}\\
&  \times\left(  e_{\mathbf{k},\sigma,j}^{(\chi)}e_{\mathbf{k}^{\prime}%
,\sigma^{\prime},j^{\prime}}^{(\chi)}e^{i\mathbf{k\cdot r}-i\omega_{k}%
t}e^{i\mathbf{k}^{\prime}\mathbf{\cdot r}^{\prime}-i\omega_{k^{\prime}%
}t^{\prime}}+e_{\mathbf{k}^{\prime},\sigma^{\prime},j}^{(\chi)}e_{\mathbf{k}%
,\sigma,j^{\prime}}^{(\chi)}e^{i\mathbf{k\cdot r}^{\prime}-i\omega
_{k}t^{\prime}}e^{i\mathbf{k}^{\prime}\mathbf{\cdot r}-i\omega_{k^{\prime}}%
t}\right) \nonumber
\end{align}
which becomes a $2$-photon wave function if we set $t^{\prime}=t$. A separate
symmetrization step is not required since its symmetric form is a direct
consequence of the commutation relations satisfied by the photon annihilation
and creation operators.

A photon density operator that counts photons by first annihilating and then
recreating a helicity $\sigma$ photon at $\mathbf{r}$, time $t$ can be defined
as%
\begin{equation}
\widehat{n}_{\sigma}^{(\alpha)}\left(  \mathbf{r},t\right)  =\widehat
{\mathbf{\psi}}_{\mathbf{r},\sigma}^{(\alpha)\dagger}\left(  t\right)
\cdot\widehat{\mathbf{\psi}}_{\mathbf{r},\sigma}^{(-\alpha)}\left(  t\right)
. \label{DensityOp}%
\end{equation}
The expectation value of this number density operator is
\begin{align}
n_{\sigma}^{(\alpha)}\left(  \mathbf{r},t\right)   &  =\left\langle
\Psi\left\vert \widehat{n}_{\sigma}\left(  \mathbf{r},t\right)  \right\vert
\Psi\right\rangle \label{Correlation}\\
&  =\sum_{j}\left\langle \Psi\left\vert \widehat{\psi}_{\mathbf{r},\sigma
,j}^{(\alpha)\dagger}\left(  t\right)  \widehat{\psi}_{\mathbf{r},\sigma
,j}^{(-\alpha)}\left(  t\right)  \right\vert \Psi\right\rangle \text{.}%
\nonumber
\end{align}
The $0$-photon contribution to $n$ is $0$, while the $1$-photon contribution
is
\begin{equation}
n_{\sigma}^{(\alpha)}\left(  \mathbf{r},t\right)  =\mathbf{\Psi}_{\sigma
}^{(\alpha)\ast}\left(  \mathbf{r},t\right)  \cdot\mathbf{\Psi}_{\sigma
}^{(-\alpha)}\left(  \mathbf{r},t\right)  . \label{PhotonDensity}%
\end{equation}
For the $2$-photon state (\ref{2Photon}), substitution of (\ref{Commutation})
gives%
\[
n_{\sigma}^{(\alpha)}\left(  \mathbf{r},t\right)  =\sum_{\sigma^{\prime
};j,j^{\prime}}\int d^{3}r^{\prime}\Psi_{\sigma,\sigma^{\prime};j,j^{\prime}%
}^{(\alpha)\ast}\left(  \mathbf{r},\mathbf{r}^{\prime};t,t\right)
\Psi_{\sigma,\sigma^{\prime};j,j^{\prime}}^{(-\alpha)}\left(  \mathbf{r}%
,\mathbf{r}^{\prime};t,t\right)  ,
\]
implying that unobserved photons are summed over. A similar argument can be
applied to each $n$-photon term. Photons are noninteracting particles and the
existence of a photon density is consistent with Feynman's conclusion the
photon probability density can be interpreted as particle density. The
$\alpha=\pm1/2$ photon density is not necessarily real or positive definite. A
real density can be obtained by averaging the $\alpha=1/2$ and $\alpha=-1/2$
densities, equivalent to taking the real part of $n_{\sigma}^{(\alpha)}\left(
\mathbf{r},t\right)  $ \cite{HawtonWaveMechanics07}. Only the $\alpha=0$
density is strictly positive definite, however.

For a negligibly small detector of volume $\Delta V$, the probability that a
photon is present in the detector is $n_{\sigma}^{(\alpha)}\left(
\mathbf{r},t\right)  \Delta V.$ To compare with Glauber photodetection theory
we can consider a pulse travelling in the $z$-direction that is normally
incident on a detector of thickness $\Delta z$ and area $\Delta A$ with
$\Delta V=\Delta A\Delta z.$ The Glauber count rate is given by (\ref{Glauber}%
) and $\left(  dn_{G}/dt\right)  \Delta z/c$ is the probability the photon
will be counted during the time that it takes to traverse the detector. Since
$n_{\sigma}^{(1/2)}=i\epsilon_{0}\mathbf{E}_{\sigma}^{(-)}\cdot\mathbf{A}%
_{\sigma}^{(+)}/\hbar$ where $A_{\sigma}^{(+)}\approx-iE_{\sigma}%
^{(+)}/\overline{\omega}$ for most beams available in the laboratory, the
predictions of the present photon number based theory and Glauber
photodetection theory are usually indistinguishable.

\section{Fields and Maxwell equations}

If we identify the operators $\widehat{\mathbf{\Psi}}^{(-1/2)}$ and
$\widehat{\mathbf{\Psi}}^{(1/2)}$ with the vector potential and displacement
operators as
\begin{align}
\widehat{\mathbf{\Psi}}_{\sigma}^{(-1/2)}\left(  \mathbf{r},t\right)   &
=\sqrt{\frac{2\epsilon_{0}}{\hbar}}\widehat{\mathbf{A}}_{\sigma}^{(+)}\left(
\mathbf{r},t\right)  ,\label{A&D}\\
\widehat{\mathbf{\Psi}}_{\sigma}^{(1/2)}\left(  \mathbf{r},t\right)   &
=-i\sqrt{\frac{2}{\hbar\epsilon_{0}}}\widehat{\mathbf{D}}_{\sigma}%
^{(+)}\left(  \mathbf{r},t\right)  ,\nonumber
\end{align}
the commutation relations (\ref{Commutation}) are in agreement with the
commutation relations satisfied by $\widehat{\mathbf{A}}_{\sigma}^{(+)}$ and
$\widehat{\mathbf{D}}_{\sigma}^{(-)}$ in QED. When one of these field
operators acts on the vacuum state as in (\ref{1PhotonEvect}), the result is a
$1$-photon position eigenvector. The magnetic induction operator can be
defined as $\mathbf{B}_{\sigma}^{(+)}=\nabla\times\mathbf{A}_{\sigma}^{(+)}$
giving%
\begin{align}
\mathbf{A}_{\sigma}^{(+)}\left(  \mathbf{r},t\right)   &  =\left\langle
0|\widehat{\mathbf{A}}_{\mathbf{r},\sigma}^{(+)}|\Psi\right\rangle
,\label{PotentiaFieldWaveFtns}\\
\mathbf{D}_{\sigma}^{(+)}\left(  \mathbf{r},t\right)   &  =\left\langle
0|\widehat{\mathbf{D}}_{\mathbf{r},\sigma}^{(+)}|\Psi\right\rangle
,\nonumber\\
\mathbf{B}_{\sigma}^{(+)}\left(  \mathbf{r},t\right)   &  =\left\langle
0|\widehat{\mathbf{B}}_{\mathbf{r},\sigma}^{(+)}|\Psi\right\rangle .\nonumber
\end{align}
These positive frequency wave functions satisfy MEs. For definite helicity
states in free space $\mathbf{D}_{\sigma}^{(+)}/\sqrt{\epsilon_{0}}%
=i\sigma\mathbf{B}_{\sigma}^{(+)}/\sqrt{\mu_{0}}=\mathbf{F}_{\sigma}%
^{(+)}/\sqrt{2}$ where $\mathbf{F}_{\sigma}^{(+)}=\mathbf{D}_{\sigma}%
^{(+)}/\sqrt{2\epsilon_{0}}+i\sigma\mathbf{B}_{\sigma}^{(+)}/\sqrt{2\mu_{0}}$
is the the Reimann-Silberstein (RS) wave function \cite{BB,ThePhoton}.
Provided we work with definite helicity states, the RS and electric or
magnetic field forms of the wave function are all equivalent.\ 

In a medium the polarization and magnetization must also be considered
\cite{HawtonWaveMechanics07} and the vacuum state can be replaced with the
zero photon state with all matter in its ground state. If the multipolar
Hamiltonian is used, the vector potential operator includes a matter part ,
$\widehat{\mathbf{A}}_{m}^{(+)}$, and the positive frequency vector potential
can be written as $\widehat{\mathbf{A}}^{(+)}\left(  \mathbf{r},t\right)
=\widehat{\mathbf{A}}_{1}^{(+)}\left(  \mathbf{r},t\right)  +\widehat
{\mathbf{A}}_{-1}^{(+)}\left(  \mathbf{r},t\right)  +\widehat{\mathbf{A}}%
_{m}^{(+)}\left(  \mathbf{r},t\right)  $. In a nonmagnetic medium,
$\widehat{\mathbf{H}}=\widehat{\mathbf{B}}/\mu_{0}$ and $\widehat{\mathbf{D}%
}=\epsilon_{0}\widehat{\mathbf{E}}+\widehat{\mathcal{P}}\ \ $where
$\widehat{\mathcal{P}}^{(+)}$ annihilates a matter excitation, analogous to
the annihilation of a photon.\ The $1$-photon MEs for the positive frequency
fields become%
\begin{align}
\nabla\cdot\mathbf{B}^{(+)}  &  =0,\ \nabla\times\mathbf{E}^{(+)}%
=-\frac{\partial\mathbf{B}^{(+)}}{\partial t},\label{MEs}\\
\nabla\cdot\mathbf{D}^{(+)}  &  =0,\ \nabla\times\mathbf{H}^{(+)}%
=\mathcal{P}\mathbf{^{(+)}+}\frac{\partial\mathbf{D}^{(+)}}{\partial
t}.\nonumber
\end{align}
These are equivalent to the equations obtained by Sipe \cite{Sipe} using an
extension of the Glauber photodetection argument.

The density $\mathbf{\Psi}_{\sigma}^{(1/2)\ast}\cdot\mathbf{\Psi}_{\sigma
}^{(-1/2)}=i\epsilon_{0}\mathbf{E}^{(-)}\cdot\mathbf{A}^{(+)}/\hbar$ has
appeared before in the classical context and in applications to beams.
Cohen-Tannoudji et. al. \cite{CT} transform the classical electromagnetic
angular momentum as
\begin{align}
\mathbf{J}  &  =\epsilon_{0}\int d^{3}r\mathbf{r\times}\left(  \mathbf{E\times
B}\right) \label{CTJ}\\
&  =\epsilon_{0}\int d^{3}r\left[  \sum_{i=1}^{3}E_{i}\left(  \mathbf{r\times
\nabla}\right)  A_{i}+\mathbf{E\times A}\right] \nonumber
\end{align}
by requiring that the fields go to zero sufficiently quickly at infinity.
Although this looks like an expectation value, the fields are classical. In a
discussion of optical beams, van Enk and Nienhuis \cite{vanEnk} separate
monochromatic fields into their positive and negative frequency parts using
\[
\mathbf{V=}\left[  \mathbf{V}^{(+)}\exp\left(  -i\omega t\right)
+\mathbf{V}^{(-)}\exp\left(  i\omega t\right)  \right]  /\sqrt{2}%
\]
and obtain for total field linear momentum and AM
\begin{align}
\mathbf{P}  &  =-i\int d^{3}r\left[  \sum_{i=1}^{3}D_{i}^{(+)\ast}\left(
i\mathbf{\nabla}\right)  A_{i}^{(+)}\right]  ,\label{vanEnkP}\\
\mathbf{J}  &  =-i\int d^{3}r\left[  \sum_{i=1}^{3}D_{i}^{(+)\ast}\left(
-\mathbf{r\times}i\mathbf{\nabla+S}\right)  A_{i}^{(+)}\right]  .
\label{vanEnkJ}%
\end{align}
Here we have assumed the absence of matter in writing $\mathbf{D=\epsilon}%
_{0}\mathbf{E}$, substituted $\mathbf{A}^{(+)}=i\omega\mathbf{D}^{(+)}$, and
changed the notation a bit for consistency with the present work. These are
classical expressions, but terms at frequency $2\omega$ do not contribute to
the total momentum and angular momentum, $\mathbf{P}$ and $\mathbf{J}$
\cite{SimmonsGuttman}. They look like the expectations values of the linear
and angular momentum operators that would be obtained using the biorthonormal
wave function pair $\sqrt{\epsilon_{0}/\hbar}\mathbf{A}^{(+)} $ and
$-i\mathbf{D}^{(+)}/\sqrt{\epsilon_{0}\hbar}$ in agreement with Eq. (\ref{A&D}).

\section{Photon wave mechanics}

Photon wave mechanics follows the usual rules, as outlined by Barut and Marlin
\cite{Barut} for example: (a) A wave equation, (\ref{WaveEquation})%
\[
i\partial\mathbf{\Psi}_{\sigma}^{(\alpha)}\left(  \mathbf{r},t\right)
/\partial t=\sigma c\nabla\times\mathbf{\Psi}_{\sigma}^{(\alpha)}\left(
\mathbf{r},t\right)  ,
\]
exists. The negative frequency solution can be eliminated on physical grounds,
thus cutting the Hilbert space in half as is done for solutions to the Dirac
equation. (b) The inner-product of the wave functions describing states
$\left\vert \widetilde{\Psi}\right\rangle $ and $\left\vert \Psi\right\rangle
,$%
\begin{equation}
\left\langle \widetilde{\Psi}^{(\alpha)}|\Psi^{(-\alpha)}\right\rangle
=\sum_{\sigma}\int d^{3}r\widetilde{\mathbf{\Psi}}_{\sigma}^{(\alpha)\ast
}\left(  \mathbf{r},t\right)  \cdot\mathbf{\Psi}_{\sigma}^{(-\alpha)}\left(
\mathbf{r},t\right)  , \label{ScalarProduct}%
\end{equation}
exists and is invariant under similarity transformations between the
$\alpha=\pm1/2\ $and the $\alpha=0$ bases. (c) The number and current
densities%
\begin{align}
n^{(\alpha)}\left(  \mathbf{r},t\right)   &  =\sum_{\sigma}\mathbf{\Psi
}_{\sigma}^{(\alpha)\ast}\cdot\mathbf{\Psi}_{\sigma}^{(-\alpha)}%
,\label{Density}\\
\mathbf{j}^{(\alpha)}\left(  \mathbf{r},t\right)   &  =-i\sigma c\sum_{\sigma
}\mathbf{\Psi}_{\sigma}^{(\alpha)\ast}\times\mathbf{\Psi}_{\sigma}^{(-\alpha
)},\nonumber
\end{align}
satisfy the continuity equation
\begin{equation}
\frac{\partial n^{(\alpha)}\left(  \mathbf{r},t\right)  }{\partial t}%
+\nabla\cdot\mathbf{j}^{(\alpha)}\left(  \mathbf{r},t\right)  =0.
\label{Continuity}%
\end{equation}
This can be verified using the wave equation. The density $n^{(0)}%
=\sum_{\sigma}\left\vert \Psi_{\sigma}^{(0)}\right\vert ^{2}$ is positive
definite, while $\left[  n^{(1/2)},\mathbf{j}^{(1/2)}\right]  $ is a
$4$-vector that can be written as the contraction of second rank EM field
tensors with $4$-potentials. (d) The momentum operator is $\hbar\mathbf{k}$
and the position operator is given by Eq. (\ref{rop}) in $\mathbf{k}$-space.
(e) The eigenvectors of the position operator are $\delta$-function
normalized. (f) The position operator and inner-product give the density
$\left\langle \mathbf{\psi}_{\mathbf{r},\sigma}^{(\alpha)}|\Psi\right\rangle
^{\ast}\left\langle \mathbf{\psi}_{\mathbf{r},\sigma}^{(-\alpha)}%
|\Psi\right\rangle $.

\section{Angular momentum and beams}

The physical interpretation of the position eigenvectors \cite{HawtonBaylisAM}
was motivated by the recent experimental and theoretical work on optical
vortices \cite{AM}, and AM is central to an understanding of position
eigenvectors. The expansion of the unit vectors $\mathbf{e}_{\mathbf{k}%
,\sigma}^{\left(  \chi\right)  }$\ in Eq. (\ref{em}) shows that the unit
vectors contribute AM $\left\{  s_{z},l_{z}\right\}  $ equal to$\ \left\{
-1,m\sigma+1\right\}  $, $\left\{  0,m\sigma\right\}  $ or $\left\{
1,m\sigma-1\right\}  $ with probability amplitudes $\left(  \cos\theta
-\sigma\right)  /2,$ $\sin\theta/\sqrt{2}$ and $\left(  \cos\theta
+\sigma\right)  /2$ respectively to a definite helicity localized state. For
every term the $z$-component of total AM is $\hbar j_{z}$ with $j_{z}%
=m\sigma,$ that is total AM has a definite value, but spin and orbital AM do
not. The position eigenvectors must have orbital AM, and this implies a vortex
structure that can be moved from to positive to the negative $z$-axis
depending on the choice of $\chi$, but not eliminated.

Theoretically, the simplest beams with orbital AM are the nondiffracting
Bessel beams (BBs), and these beams are closely related to our localized
states. BBs satisfy MEs and have definite frequency, $ck_{0},$ and a definite
wave vector, $k_{z},$ along the propagation direction$.$ It then follows that
the $\mathbf{k}$-space transverse wave vector magnitude $k_{\bot}=\sqrt
{k_{0}^{2}-k_{z}^{2}},$ and the angle $\theta=\tan^{-1}\left(  k_{\bot}%
/k_{z}\right)  $ also have definite values. Cylindrical symmetry is retained
when integrating over $\phi$ by weighting all $\phi$ equally with phase factor
$\exp\left(  im\phi\right)  .$ When Fourier transformed to $\mathbf{r}$-space
the modes go as $\exp\left(  -ik_{0}ct+il_{z}\varphi+ik_{z}z\right)  J_{l_{z}%
}\left(  k_{\perp}r\right)  $ where $l_{z}=m$ and $m\pm1$ in (\ref{em}),
$J_{l_{z}}$ are Bessel functions, $\varphi$ the real space azimuthal angle,
and $r$ is the perpendicular distance from the beam axis \cite{Hacyan}. If
integrated over $k_{\bot}$ the result is a sum of outgoing and incoming waves
that is localized on the $z$-axis at some instant in time. If the BBs are then
integrated over $k_{z},$ the result is equivalent to a sum over all positive
and negative wave vectors, and states localized in three dimensions are
obtained. If we select $\chi=0$ so that the $\mathbf{k}$-space unit vectors
are $\widehat{\mathbf{\phi}}$ and $\widehat{\mathbf{\theta}}$ in the linear
polarization basis, $\mathbf{B}$ is transverse to $\widehat{\mathbf{z}}$ for
the $\widehat{\mathbf{\theta}}$ mode and $\mathbf{E}$ is transverse for the
$\widehat{\mathbf{\phi}}$ mode and the linearly polarized modes can be called
transverse magnetic (TM) and transverse electric (TE) respectively.

A paraxial beam propagating in the $\widehat{\mathbf{z}}$-direction with
frequency $\omega,$ helicity $\sigma,$ and $z$-component of orbital AM $\hbar
l_{z}$ can be described in cylindrical polar coordinates by the vector
potential \cite{Allen92}%
\begin{equation}
\mathbf{A}^{(+)}\left(  \mathbf{r},t\right)  =\frac{1}{2}\left(
\widehat{\mathbf{x}}+i\sigma\widehat{\mathbf{y}}\right)  u\left(  r\right)
\exp\left[  il_{z}\varphi+ik_{z}\left(  z-ct\right)  \right]  . \label{Abeam}%
\end{equation}
This vector potential is equivalent to the wave function $\mathbf{\Psi
}_{\sigma^{\prime}}\left(  \mathbf{r},t\right)  =\delta_{\sigma,\sigma
^{\prime}}\sqrt{2\epsilon_{0}/\hbar}\mathbf{A}^{(+)}\left(  \mathbf{r}%
,t\right)  $. The $z$-component of the time average of the classical AM
density, equal to $\mathbf{r\times}\left(  \mathbf{D}^{(-)}\mathbf{\times
B}^{(+)}\right)  ,$ is then found to be%
\begin{equation}
J_{z}\left(  r\right)  =\epsilon_{0}\left[  \omega l_{z}\left\vert u\left(
r\right)  \right\vert ^{2}-\frac{1}{2}\omega\sigma r\frac{\partial\left\vert
u^{2}\left(  r\right)  \right\vert }{\partial r}\right]  . \label{Jbeam}%
\end{equation}
This can be interpreted as the quantum mechanical AM density in a $1$-photon
state. The first term of (\ref{Jbeam}) is consistent with orbital AM $\hbar
l_{z}$ per photon since the photon density given by (\ref{PhotonDensity})
reduces to $n^{(1/2)}\left(  \mathbf{r},t\right)  =\epsilon_{0}\omega
\left\vert u\left(  r\right)  \right\vert ^{2}/\hbar.$ The last term of
(\ref{Jbeam}) does not look like photon spin density. The most paradoxical
case is a plane wave. For example a wave function proportional to $\left(
\widehat{\mathbf{x}}+i\sigma\widehat{\mathbf{y}}\right)  \exp\left(
ikz-i\omega t\right)  $ implies linear momentum $\hbar k\widehat{\mathbf{z}}$
per photon\ and hence no $z$-component of AM. The AM of this beam resides in
its edges, as can be seen from Eq. (\ref{Jbeam}) and a new edge is created if
the disk intercepts part of the beam, reducing the AM of the beam
\cite{AllenPadgett}. Based on the derivation here, this result applies not
only to a classical beam, but also to a $1$-photon state.

\section{Summary}

We have derived $1$ and $2$-photon wave functions from QED by projecting the
state vector onto the eigenvectors of a photon position operator. Largely
because it is still widely believed that there is no position operator, this
is the first time that photon wave functions have been obtained in this way.
While only the LP wave function gives a positive definite photon density,
field-like wave functions are widely used and are more convenient in many
applications. In the field-like definite helicity basis the wave function pair
is given by (\ref{A&D}), with $\mathbf{\Psi}_{\sigma}^{(\alpha)}$ are given by
(\ref{1PhotonPsi}). The linear polarization basis of TM and TE fields can be
obtained by taking the sum and difference of the definite helicity modes as in
(\ref{rBasis}). The $1$-photon density is $n_{\sigma}^{(\alpha)}\left(
\mathbf{r},t\right)  =\mathbf{\Psi}_{\sigma}^{(\alpha)\ast}\cdot\mathbf{\Psi
}_{\sigma}^{(-\alpha)},$where $n_{\sigma}^{(1/2)}$ is essentially equal to
$n_{\sigma}^{(0)}$except for very broad band signals. This can be generalized
to describe the photon density in a multiphoton state using the expectation
value of the number operator, (\ref{Correlation}).

Systematic investigation of photon position operators and their eigenvectors
clarifies the role of the photon wave function in classical and quantum
optics. The LP wave function is related to field based wave functions through
a similarity transformation that preserves eigenvalues and scalar products.
The field $\mathbf{D}^{(+)}\left(  \mathbf{r},t\right)  $ is proportional to
the Glauber wave function which gives the photodetection amplitude for a
detector that responds to the electric field. Fields and potentials are
locally related to charge and current sources, and hence are the most
convenient in many applications. However, Fourier transformation of
$\mathbf{k}$-space probability amplitudes naturally leads to the LP form
\cite{Eberly}. Selection of the LP or field-potential is a matter of
convenience in most applications.

By the general rules of quantum mechanics the LP wave function gives the
probability to detect a photon at a point in space. It and the closely related
field-potential wave function pair obtained by solution of MEs are ideally
suited to the interpretation of photon counting experiments using a detector
that is small in comparison with the spatial variations of photon density. It
is not subject to limitations based on nonlocalizability, and coarse graining
or restriction to length scales smaller than a wave length is not required.
Exact localization in vacuum requires infinite energy and is not physically
possible, but position eigenvectors provide a useful mathematical description
of photon density. Photon number density is equivalent to integration over
undetected photons in a multiphoton beam. Our formalism justifies the use of
positive frequency solutions to MEs as photon wave functions and gives a
rigorous theoretical basis for extrapolation of their range of applicability
from the many photon to the $1$-photon regime. \ 

\acknowledgments

The author acknowledges the financial support of the Natural Science and
Engineering Research Council of Canada.

\bibliographystyle{spiebib}
\bibliography{report}

\newpage

\end{document}